\documentclass[twocolumn, tighten, dvipsnames, floatfix]{aastex631}

\usepackage{soul}

\hypersetup{linkcolor=red,citecolor=green,filecolor=cyan,urlcolor=magenta}
\hypersetup{
    colorlinks=true,
    linkcolor=blue,
    filecolor=magenta,
    anchorcolor=red,      
    urlcolor=purple,
    citecolor=blue,
    }

\received{}
\revised{}
\accepted{}
\submitjournal{ApJL}

\shorttitle{SOS1 star from Terzan 5}
\shortauthors{Souza et al.}

\begin{document}

\title{Tracing back a second-generation star stripped from Terzan 5 by the Galactic bar}

\correspondingauthor{Stefano O. Souza}
\email{ssouza@aip.de, s-souza@mpia.de, stefano.o.souza.0411@gmail.com}

\author[0000-0001-8052-969X]{Stefano O. Souza}
\affiliation{Leibniz-Institut für Astrophysik Potsdam (AIP), An der Sternwarte 16, Potsdam, D-14482, Germany}
\affiliation{Universidade de S\~ao Paulo, IAG, Rua do Mat\~ao 1226, Cidade Universit\'aria, S\~ao Paulo 05508-900, Brazil}
\affiliation{Max Planck Institute for Astronomy, K\"onigstuhl 17, D-69117 Heidelberg, Germany}

\author[0000-0003-0974-4148]{Marica Valentini}
\affiliation{Leibniz-Institut für Astrophysik Potsdam (AIP), An der Sternwarte 16, Potsdam, D-14482, Germany}

\author[0000-0003-1269-7282]{Cristina Chiappini}
\affiliation{Leibniz-Institut für Astrophysik Potsdam (AIP), An der Sternwarte 16, Potsdam, D-14482, Germany}

\author[0000-0002-5974-3998]{Angeles P\'erez-Villegas}
\affiliation{Instituto de Astronom\'ia, Universidad Nacional Aut\'onoma de M\'exico, A. P. 106, C.P. 22800, Ensenada, B. C., M éxico}

\author[0000-0002-9143-9988]{Josefina Montalb\'an}
\affiliation{Department of Physics \& Astronomy, University of Bologna, Via Gobetti 93/2, 40129, Bologna, Italy}
\affiliation{School of Physics and Astronomy, University of Birmingham, Edgbaston, Birmingham B15 2TT, UK}

\author[0000-0002-9480-8400]{Diego Bossini}
\affiliation{Instituto de Astrof\'isica e Ci\^encias do Espa\c co, Universidade do Porto, CAUP, Rua das Estrelas, 4150-762, Porto, Portugal}
\affiliation{Dipartimento di Fisica e Astronomia Galileo Galilei, Universit\`a di Padova, Vicolo dell'Osservatorio 3, 35122, Padova, Italy}
\affiliation{Osservatorio Astronomico di Padova - INAF, Vicolo dell'Osservatorio 5, 35122, Padova, Italy}

\author[0000-0001-9264-4417]{Beatriz Barbuy}
\affiliation{Universidade de S\~ao Paulo, IAG, Rua do Mat\~ao 1226, Cidade Universit\'aria, S\~ao Paulo 05508-900, Brazil}

\author{Yvonne Elsworth}
\affiliation{School of Physics and Astronomy, University of Birmingham, Edgbaston, Birmingham B15 2TT, UK}

\author[0000-0002-8854-3776]{Rafael A. Garcia}
\affiliation{Universit\'e Paris-Saclay, Universit\'e Paris Cit\'e, CEA, CNRS, AIM, F-91191 Gif-sur-Yvette, France}

\begin{abstract}
The Galactic bulge hosts the Milky Way's oldest stars, possibly coming from disrupted globular clusters (GCs) or the bulge's primordial building blocks, making these stars witnesses to the Galaxy's early chemical enrichment. The Galactic bar currently dominates the bulge's region, altering the orbits of objects formed before its formation and complicating the trace of the field stars' original clusters. Here, we present the discovery of a fossil record of this evolution, SOS1—a star trapped in the bar, exhibiting significant enhancements in nitrogen, sodium, and aluminum, typical of second-generation GC stars. SOS1 also shows an s-process Ce enhancement, suggesting an old age and early enrichment by fast-rotating massive stars in the Galaxy's earliest phases. With the purpose of finding the SOS1's parent GC, we derive its precise chemodynamical properties by combining high-precision proper motions from Gaia with APOGEE detailed chemical abundances. Our analysis {suggests that SOS1 was possibly} stripped from the GC Terzan 5 by the Galactic bar's gravitational influence approximately 350 Myr ago. {We also found chemical similarities suggesting that SOS1 belonged to the most metal-poor, ancient, and peripheral stellar population of Terzan 5. These results not only} support the hypothesis that Terzan 5 is a remnant of a primordial building block of the Galactic bulge, {but also suggest this cluster continues losing stars to the bar. Our method highlights how powerful the use of chemodynamical properties in the Gaia era is for tracing the Galaxy's evolutionary history.}
\end{abstract}

\keywords{Galactic bar(2365) --- Galactic archaeology(2178) --- Globular star clusters(656) --- Galaxy abundances(574) --- Milky Way dynamics(1051) }

\section{Introduction} \label{sec:intro}

The inner Galaxy is the most complex component of the Milky Way (MW), including a mix of an early classical bulge, inner halo, inner disk, a bar-induced pseudobulge, as well as debris of past accretion events \citep{perezvillegas20,queiroz21,horta21}. It is in the Galactic bulge that the oldest MW globular clusters (GCs) are found \citep{barbuy_review,souza23,bica24}. Among them, Terzan 5, Liller 1, and UKS~1 were suggested as representatives of early building blocks, also called Bulge Fossil Fragments \citep[BFFs;][]{ferraro09,ferraro16,crociati23}. These massive GCs are good candidates for being the local counterparts of the large star formation clumps being observed at high redshift \citep[e.g.][]{debattista23} and could be tracers of the earliest chemical enrichment phases of the Universe.

The Galactic bar should form somewhat later \citep[$3-8$ Gyr;][]{bovy19,wylie22,nepal23} through the gravitational instability of the disk possibly induced by merger events \citep{merrow24}. Stars on bar-shaped orbits span different stellar populations, collecting material from the outer to the inner regions, forming a box-shaped structure as a result of the buckling instability of the bar. The combination of Apache Point Observatory Galactic Evolution Experiment data release 17 \citep[APOGEE DR17;][]{apogee17} abundances and \textit{Gaia} \citep{gaia18} data has made it possible to trace the chemo-dynamical properties of stars in bar shape orbits confirming that most of them are metal-rich stars (and mostly low [$\alpha$/Fe]). Additionally, a non-negligible fraction was found to be metal-poor \citep{queiroz21}, with around 1/10th the solar metallicity, and had high [$\alpha$/Fe] ratios, similar to bulge GCs metallicities and alpha-enhancement. 

Along the bar formation processes, pre-existing bulge and disk material can be captured by the bar,  accelerating the evaporation of stars from previously existing GCs and oldest building blocks such as BFFs. Identifying those field stars as evaporated from a specific GC or BFF is challenging because their chemistry can be similar to most field stars. The exception is N- and Al-rich stars, whose chemical pattern is well explained considering them as second-generation (2G) stars that originate in GCs \citep[e.g.][]{carretta09b,gratton12}. Therefore, these stars are good tracers of GCs debris \citep[e.g.][]{fernandeztrincado19,fernandez-trincado22,horta21}. 

Terzan 5 is one of the most massive GCs in our Galaxy \citep[$2\times 10^6$ M$_\odot$;][]{Lanzoni10} {located in the inner region \cite[d$_{\rm GC} = 1.65 \pm 0.13$ kpc;][]{baumgardt22} } obscured by a high reddening \citep[$E(B-V)=2.38$;][]{valenti07}. It is assumed to be originally the nuclear star cluster (NSC) of a building block of the early Galactic bulge \citep{ferraro09,ferraro16}. Differently from $\omega$-Centauri which preserve characteristics of NSC \citep{neumayer20,limberg22,max24}, predictions suggest that Terzan 5 has lost $\sim 60\%$ of its initial mass \citep{romano23} since the formation of the Galactic bulge. Nevertheless, the evidence of NSC formation via merger of clusters spiralling to the center is still imprinted in Terzan 5 three-modal metallicity distribution function \cite[MDF;][]{origlia11,origlia13}: [Fe/H]$=+0.30$ (popA), [Fe/H]$=-0.30$ (popB), and [Fe/H]$=-0.87$ (popC). These characteristics indicate that Terzan 5 should have contributed with many of the stars located in the Galactic center field population. In this work, we analyze the influence of the BFF, specifically Terzan 5, on the formation of the Galactic bar through its contribution to N-Al-rich stars to the field stellar population. 

This letter is organized as follows. Section \ref{sec:sos1} presents the star analysed in this work as well as its important chemical and orbital information. In Section \ref{sec:parent_cluster}, we show the methodology developed to find the most probable parent clusters of our target. Section \ref{sec:results} shows the further comparison between our target and Terzan 5 in terms of dynamics and chemical abundances. Finally, the conclusions are drawn in Section \ref{sec:conclusions}.

\section{Data}\label{sec:sos1}

\begin{figure*}
    \centering
    \includegraphics[width=0.95\textwidth]{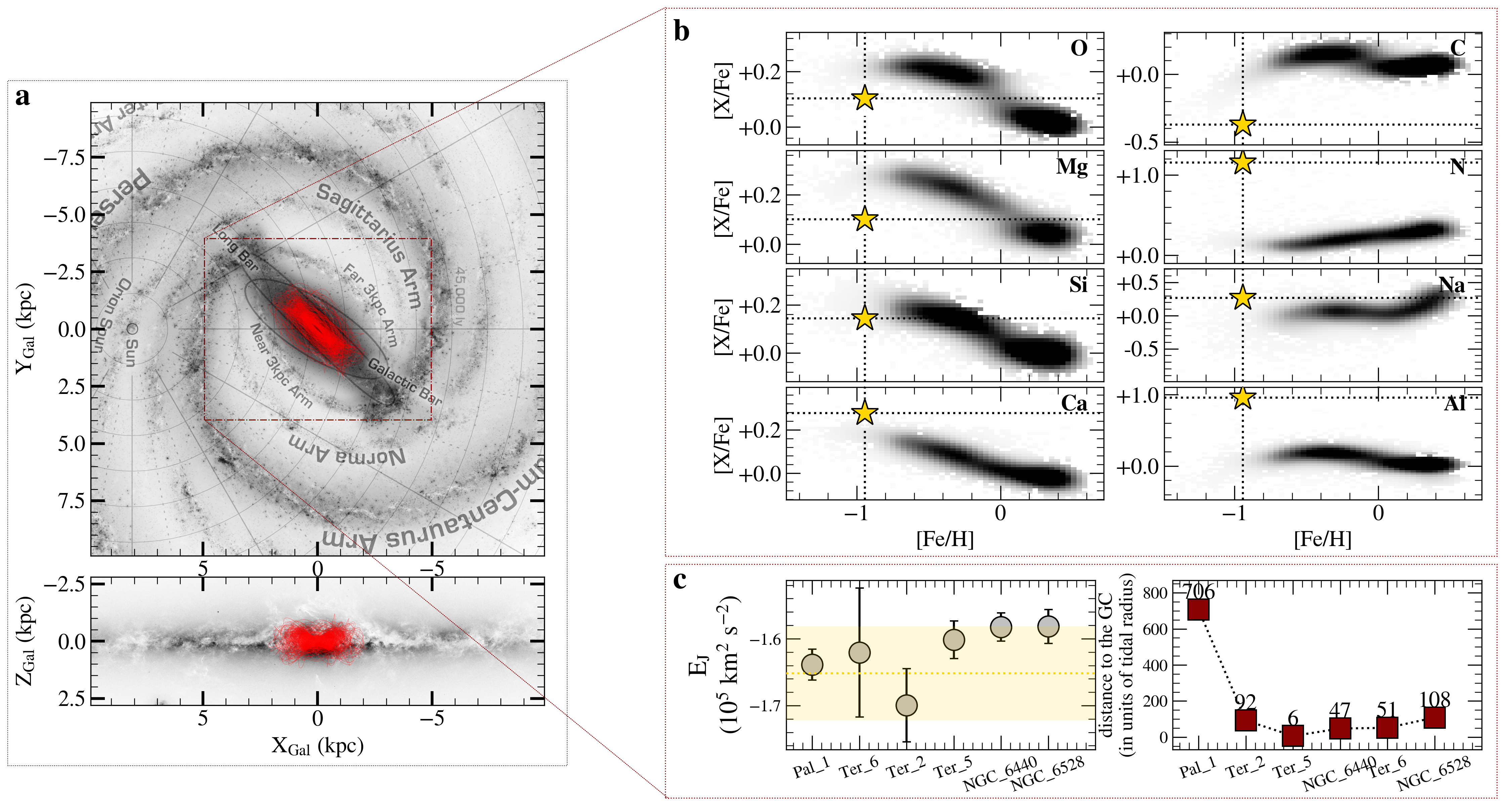}
    \caption{{Chemo-dynamical properties of SOS1}. {a}, X-Y and X-Z density projections of the Galaxy with the orbit of SOS1 star (red line). {b}, abundances of and light-elements for SOS1 (yellow star symbol) compared with field stars in the inner Galaxy (density plot). {c}, the left panel shows the Jacobi energy of the GCs compatible within $1\sigma$ with SOS1 (shaded yellow region), and the right panel shows the SOS1 distance to the centre of the selected clusters. }
    \label{fig:main_plot}
\end{figure*}

The star 2M17454705-2639109 (hereafter SOS1) analyzed in this work is located in the inner region of the MW (d$_{\rm GC}=1.43\pm0.72$ kpc and $A_V=4.48\pm0.91$ -- Apendix \ref{app:dibs}) at Galactic coordinates $l=1.97^\circ$ and $b=1.16^\circ$, at a heliocentric distance of $6.9\pm1.0$ kpc. Its orbit exhibits a bar-shaped trajectory, indicating that the star is currently trapped in the MW bar (red line in panel a of Figure \ref{fig:main_plot}). 

SOS1 is part of the sample observed with the NASA-\emph{Kepler} project $K2GO4_2-0125$ (PI: M. Valentini) in the K2 Campaign 11, targeting the Galactic bulge/thick disk region. We crossmatched the K2 sample with both \textit{Gaia} DR3 and APOGEE DR17, obtaining proper motions, radial velocities, chemical abundances, and atmospheric parameters. The information is listed in Table \ref{tab:informations}. 

The APOGEE abundances and atmospheric parameters were derived using the APOGEE Stellar Parameter and Chemical Abundances Pipeline \citep[ASPCAP;][]{garcia-perez16}. Since the errors provided by the pipeline are the internal errors from the method employed, they can be underestimated. Therefore, in Table \ref{tab:informations} we added to the errors in effective temperature and chemical abundance ratios $50$K and $0.05$ dex, respectively. SOS1 has SNR$>50$, RUWE (renormalised unit weight error) $<1.4$, \texttt{APSCAPFLAGs}$=0$, \texttt{STARFLAG}$=0$, and \texttt{FLAG}$=0$ for all elements. These quality cuts are similar and recommended in recent studies using APOGEE DR17 data \citep[e.g.][]{horta21,queiroz21,limberg22,souza23}.

\section{SOS1 parent cluster}\label{sec:parent_cluster}
In this section, we present the clues for the most likely SOS1 parent cluster.

\subsection{Chemical clue}
SOS1 exhibits an abundance pattern (Figure \ref{fig:main_plot}b) of $\alpha$- and light-elements different than that of field stars in the innermost regions of the Galaxy \citep{queiroz21}. From the APOGEE DR17 values, SOS1 presents high nitrogen (N) and aluminum (Al) abundances ([N/Fe]$>1.0$, [Al/Fe]$>1.0$) and a carbon (C) depletion ([C/Fe]$<$-0.2), which are signatures of 2G GC stars \citep[e.g.][]{gratton12,meszaros20}. To validate the abundances provided by APOGEE, we performed two different sanity checks to confirm the high values of N and Al abundances. The first test compares the SOS1 spectrum to the spectrum of a reference star (STARB in Table \ref{tab:informations}) with similar atmospheric parameters, with, however, low N and Al abundances; in the second test we roughly measured the abundances through a line profile fitting using synthetic spectra. The details of the validation test are given in the Appendix \ref{app:abds_verify}. Therefore, given its orbit and chemistry, SOS1 is a good candidate for an evaporated star from a GC currently trapped by the bar.

Two main scenarios could be evoked to explain SOS1. In the first scenario, the star could have been ejected from the cluster due to a strong encounter (e.g., star-binary interaction). {While such encounters are capable of producing hypervelocity stars with velocities $>10^3$ km s$^{-1}$, the typical ejection speed is around 30\% higher than the escape velocity of the cluster's core \citep{weatherford23}. For massive GCs, the escape velocity is $\sim 50$ km s$^{-1}$ \citep{baumgardt18}}. In the second scenario, GCs that are tidally interacting with the Galactic bar can lose stars to the Galaxy via tidal stripping, mainly affecting stars located in the outer parts of the cluster \citep{moreno22}. The star is then slowly stripped from the cluster during several passages of the GC through the bar. {Even though SOS1's total velocity with respect to the Galactic center of $96\pm3$ km s$^{-1}$ could still be consistent with an ejection from a strong encounter,  more than 75\% of stars escape from GCs through a combination of weak encounters and tidal effects \citep{weatherford23}, being the most plausible scenario for SOS1}. Recently, \cite{minniti2024} found a GC that is falling towards the center of the MW, illustrating a process that could explain SOS1 and likely occurred in many clusters throughout the history of the Galaxy.

SOS1 is overabundant in \emph{s}-process element Cerium ([Ce/Fe] = +0.43 - see Table \ref{tab:informations}), mainly produced by low- and intermediate-mass stars and restored to the interstellar medium on longer timescales. Fast-rotating massive stars can, however, boost the production of \emph{s}-process elements \citep{frischknecht16} and explain large enhancements of Ce early on in the Universe. In addition, the fast chemical enrichment expected to have occurred in the innermost regions of the MW can explain the moderate metallicities of the oldest bulge GCs and field stars \citep{chiappini11,barbuy_review,souza24}. Other recent studies have found Ce-enhanced stars in the bulge, similar to what we find here for SOS1 \citep[e.g.][]{razera22,sales-silva24}. This chemical property is also compatible with SOS1 being a fossil record of the earliest chemical enrichment phases of the MW when fast-rotating massive stars would have played an important role.

\begin{table}[hbt!]
\centering
\caption{ \textbf{Radial velocity, astrometric, atmospheric, and chemical information of SOS1 and STARB stars}}
\label{tab:informations}
\resizebox{\columnwidth}{!}{%
\begin{tabular}{l|cc}
    \hline
     & SOS1  &  STARB \\
     APOGEE ID & 2M17454705-2639109 & 2M07273047-7511343 \\
     \hline
     RV (km\,s$^{-1}$)             & $-75.41\pm0.04$ & $+63.48\pm0.04$ \\
     $\mu_{\alpha}^*$ (mas yr$^{-1}$) & $-1.45\pm0.10$ & $+5.12\pm0.01$ \\
     $\mu_{\delta}$ (mas yr$^{-1}$)   & $-6.20\pm0.06$ & $-4.72\pm0.01$ \\
     T$_{\rm eff}$ (K)                & $4366\pm61$    & $4357\pm62$  \\
     $\log g$                         & $1.51\pm0.04$  & $1.51\pm0.04$  \\
     SNR                              & $129$          & $117$  \\
     RUWE                             & $1.008$        & $1.078$  \\
     \texttt{ASPCAPFLAG}              & $0$            & $0$  \\
     \texttt{STARFLAG}                & $0$            & $0$  \\
      d$_\odot$ (kpc)                 & $6.86\pm0.81$  & $6.15\pm0.21$  \\
      A$_V$ (mag)                     & $4.48\pm0.92$  & $0.87\pm0.18$  \\
     \hline
     $[$Fe/H$]$                &    $-0.94\pm0.06$ &     $-1.07\pm0.06$  \\
     $[$C/Fe$]$                &    $-0.37\pm0.08$ &     $-0.36\pm0.08$  \\
     $[$N/Fe$]$                &    $+1.15\pm0.08$ &     $+0.11\pm0.08$  \\
     $[$C/N$]_{\rm corrected}$ &    $+0.18\pm0.08$ &                 --  \\
     $[$O/Fe$]$                &    $+0.13\pm0.08$ &     $+0.28\pm0.08$  \\
     $[$Mg/Fe$]$               &    $+0.13\pm0.07$ &     $+0.19\pm0.07$  \\
     $[$Ca/Fe$]$               &    $+0.35\pm0.08$ &     $+0.14\pm0.09$  \\
     $[$Si/Fe$]$               &    $+0.18\pm0.07$ &     $+0.17\pm0.08$  \\
     $[$Na/Fe$]$               &    $+0.27\pm0.14$ &     $+0.32\pm0.18$  \\
     $[$Al/Fe$]$               &    $+0.96\pm0.08$ &     $-0.16\pm0.09$  \\
     $[$Mn/Fe$]$               &    $-0.17\pm0.08$ &     $-0.34\pm0.09$  \\
     $[$V/Fe$]$                &    $+0.14\pm0.13$ &     $+0.16\pm0.09$  \\
     $[$Ce/Fe$]$               &    $+0.43\pm0.08$ &     $-0.18\pm0.08$  \\
     $[$Co/Fe$]$               &    $+0.23\pm0.11$ &     $-0.78\pm0.13$  \\
    \hline
\end{tabular}}
\end{table}

\subsection{Dynamical clue}

A star captured by the bar is expected to retain a) the chemical composition of the material from which it was formed, and b) the original dynamical signature as the Jacobi energy (E$_{\rm J}$) for a longer time.  In what follows, we use both dynamical and chemical information to identify the most probable GC from which SOS1 originally belonged. 

In this work, we performed the orbital analysis employing the analytic approximation by \cite{sormani22} of the $N-$body Galactic model from \cite{portail17} constructed to be used in combination with the Action-based GAlaxy Modelling Architecture \citep[AGAMA;][]{agama} Python code. We integrated the orbits 13 Gyr backwards. Since our focus relies on the Galactic centre, this model is more convenient since it fits well with the kinematics and stellar density distribution of the inner Galaxy, including the bar and X-shape structures. As a consequence of the adopted Galactic potential, which includes an axisymmetric background and a bar component that rotates, there is no conservation of the total energy (E$_T$) and the z-component of the angular momentum (L$_Z$) along the orbit integration. Instead of these quantities, in the bar rotating frame, the E$_J$ is conserved. The E$_J$, in units of mass, is calculated as follows:

\begin{equation}
    \displaystyle E_J = \frac{1}{2} \mathbf{V'}^2 + \Phi(\mathbf{r'}) - \frac{1}{2} \Omega^2_p \left( x^{'2} + y^{'2}\right)
\end{equation}
where $\Omega_p$ is the pattern speed of the bar, assumed to be $-39$ km s$^{-1}$ kpc$^{-1}$ \citep{portail17}.

Assuming that the parent cluster of SOS1 is still orbiting the MW, we selected the GCs with E$_{\rm J}$ compatible within $1\sigma$ with the E$_{\rm J}$ of SOS1 (Figure \ref{fig:main_plot}c left panel), this sample consisting of the most probable SOS1's parent clusters (see all their information in Table \ref{tab:gcs_ref}). The closest GC to SOS1 with similar E$_{\rm J}$ (as selected before) is Terzan 5 (Figure \ref{fig:main_plot}c right). At the same time, SOS1 is far from the Terzan 5's center by six times its tidal radius, showing that SOS1 is no longer gravitationally bound to any cluster.

\begin{figure}
    \centering
    \includegraphics[width=\columnwidth]{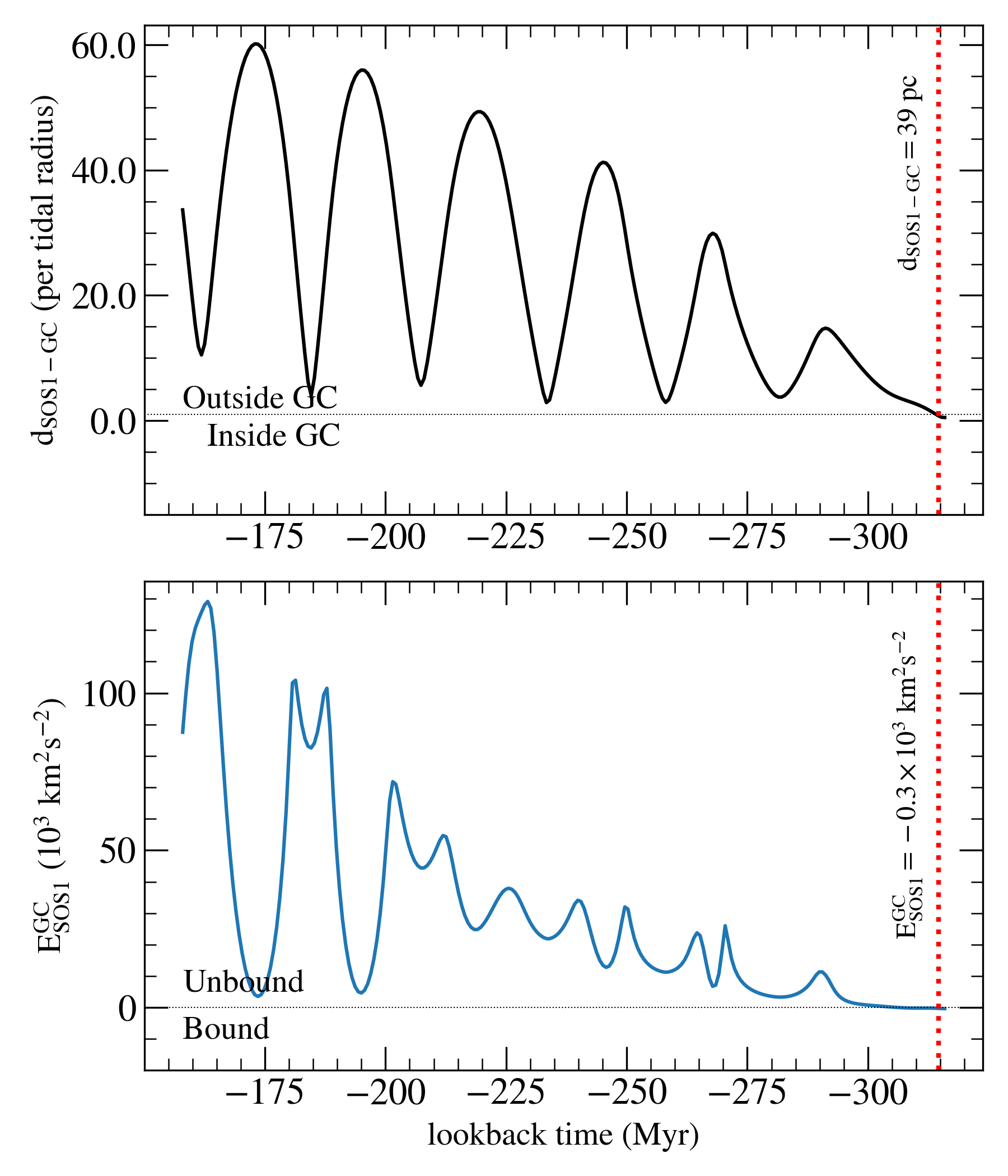}
    \caption{\textbf{Dissociation point selection criteria \href{https://stefanoos.github.io/sos1_files//interactive_figures/dissociation_time_new.html}{(Interactive figure)}}. The top panel shows the SOS1 distance to GC's centre in units of tidal radius as a function of time. The black horizontal dotted line shows the tidal radius region limiting the cluster volume. The bottom panel shows the SOS1 total energy with respect to the GC as a function of time. The black horizontal dotted line represents the bound energy limit. The red dotted lines show the dissociation point with the corresponding distance to the cluster centre (top panel) and the bound energy (bottom panel).}
    \label{fig:orbits}
\end{figure}

To find the most probable former GC of SOS1, we compared its orbit with that of individual GCs (Appendix \ref{app:orbits} provides more details on the orbital integration). For each pair SOS1-GC, a total of $2.5\times10^{5}$ combinations of orbits are analysed, and we searched for a crossing point in each one. Nevertheless, if the star has a velocity high enough at the crossing point, it can pass through the cluster volume without being bound to it. Therefore, to consider the crossing as a dissociation point\footnote{The \emph{dissociation point} is defined as the last orbital point when the star was finally captured from a specific cluster by the Galactic potential.}, the star must be within the cluster volume defined as a tidal radius and gravitationally bound to the cluster (Figure \ref{fig:orbits}). 

By performing the dissociation point analysis for each pair, Terzan 5 was determined to be the most probable parent cluster. Although the number of found dissociation points for SOS1-Terzan 5 is a small fraction of the number of orbital tracebacks performed, it is still larger than the number of dissociation points found between SOS1 and any of the other GCs by a factor ${>}200$ (see Table \ref{tab:gcs_ref}). This is especially true when limiting our analysis to only the past 2 Gyr, during which the employed Galactic potential should be most accurate; in this time, there are ${>}900$ dissociation points between SOS1 and Terzan 5, but none between SOS1 and any other observed GC (see Table \ref{tab:gcs_ref}). This dynamical finding suggests that Terzan 5 is more likely to have spawned SOS1 than any other observed GC.

It is worth mentioning the fact that our boundedness criterion for identifying a dissociation point is an approximation, which assumes that the GC is not evolving (e.g. fixed tidal radius) and the orbits are only under the influence of the Galactic potential with no dynamical friction effect.

\section{Terzan 5 as the SOS1 parent cluster}\label{sec:results}

\begin{figure}
    \centering
    \includegraphics[width=\columnwidth]{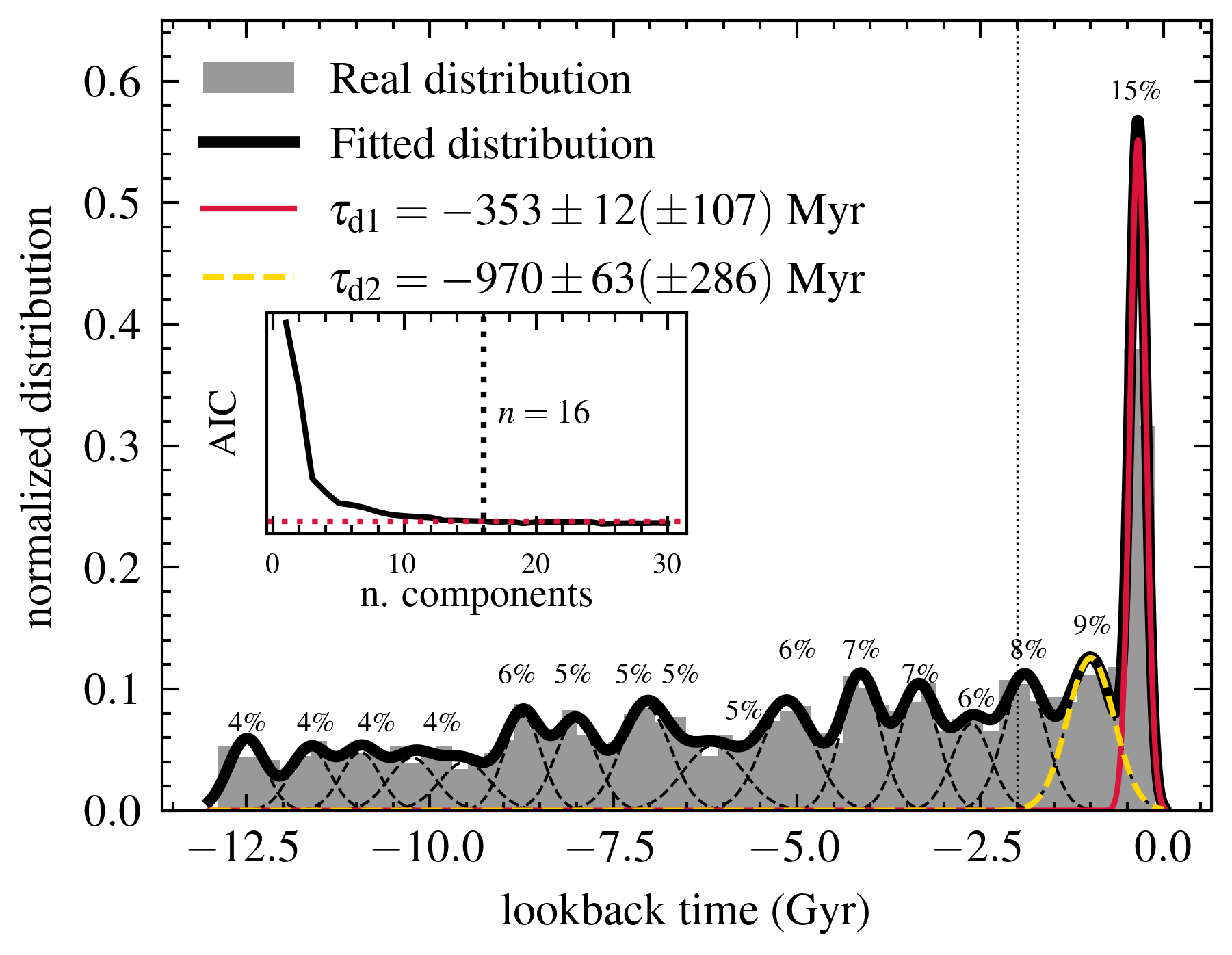}    \caption{\textbf{Dissociation time calculation}. The gray histogram shows the distribution of lookback time obtained from the dissociation points. The black solid line shows the multiple Gaussian distribution obtained through the GMM fitting, and the individual Gaussians are as black thin dashed lines. The two significant dissociation times $\tau_{\rm d1}$ and $\tau_{\rm d2}$ are highlighted as red and yellow lines, respectively. The number above each Gaussian represents the individual contribution to the total fitting. The inset plot shows the AIC test.}
    \label{fig:dissociation-time_result}
\end{figure}

Recovering the time at the dissociation point, which we call the dissociation time ($\tau_d$), the result is a time distribution where we applied the Gaussian Mixture Models (GMM) fitting to obtain the time when the bar captured the star from the cluster. To avoid overfitting the distribution, we determined the number of individual Gaussian components to use in the GMM by using the Akaike information criterion (AIC), specifically choosing the first number of components for which the AIC parameter becomes constant (insert plot in Figure \ref{fig:dissociation-time_result}). For Terzan 5, we found 16 dissociation time distributions employing the AIC test (Figure \ref{fig:dissociation-time_result}). For the interval between 0 and -2 Gyr, where we can trust our analysis, we found two dissociation times $\tau_{\rm d1} = -353\pm12(\pm107)$ Myr (red line in Figure \ref{fig:dissociation-time_result} ) and $\tau_{\rm d2} = -970\pm63(\pm286)$ Myr (yellow line). The $\tau_{\rm d1}$ distribution contributes $15\%$ to the total distribution (black line), while $\tau_{\rm d2}$ $9\%$. We performed 2000 Monte Carlo realizations to obtain the uncertainties, and we are providing two values. The first one represents the uncertainty in the mean value, while the second one is the standard deviation of the distribution. Even though the time distribution presents values up to $13$ Gyr ago, these points do not make physical sense because the Galaxy potential should evolve with time, and at that time, even the bar had not been formed yet. Therefore, the results for SOS1-Terzan 5 are within the time range having a physical meaning. In summary, the dynamical results suggest that SOS1 was possibly gravitationally bound to Terzan 5 at $-353\pm12(\pm107)$ Myr ago. 

\begin{figure*}
    \centering
    \includegraphics[width=\textwidth]{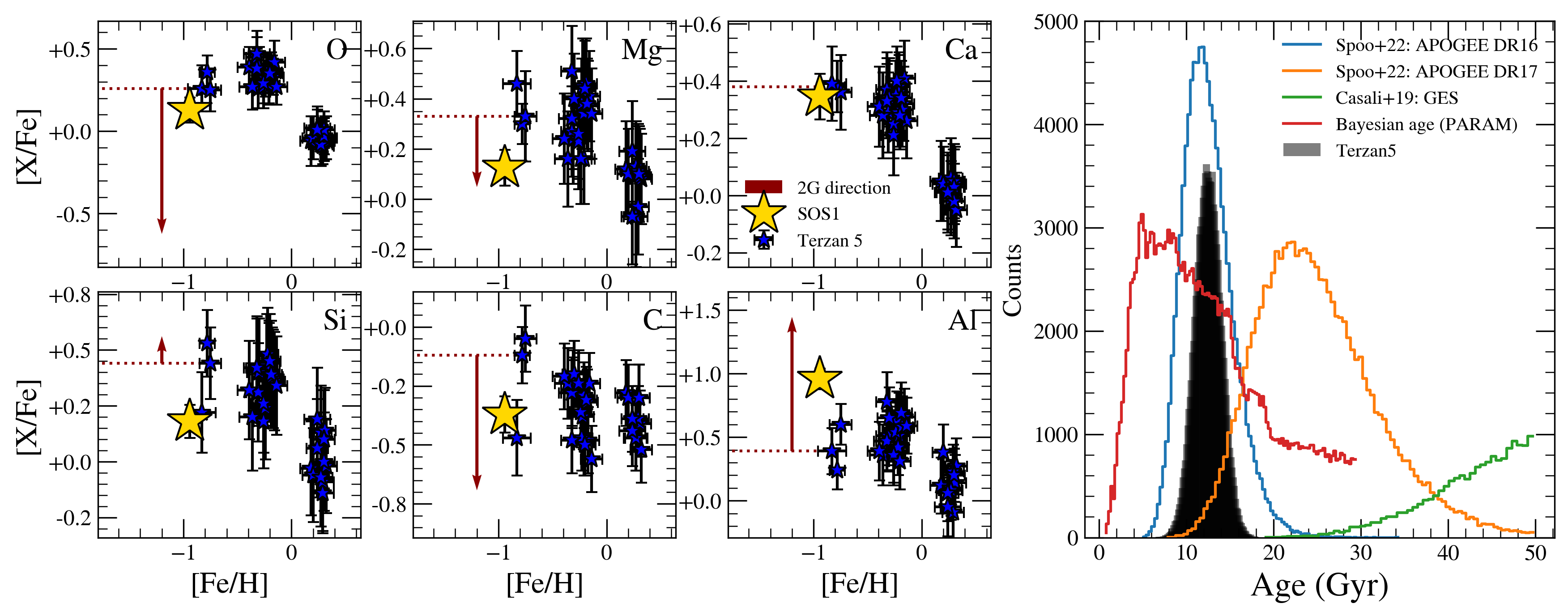}
    \caption{ \textbf{Chrono-Chemical comparison between SOS1 and Terzan 5}.
    The abundances of $\alpha$ elements, C, and Al vs. the iron content [Fe/H] for Terzan 5 \protect\citep{origlia11,origlia13} (blue) and the SOS1 (yellow). The arrow in each panel shows the average maximum difference \protect\citep{milone18} between 1G (dotted line) and 2G for a metallicity around [Fe/H]$= -1.0$, the Terzan 5's popC (the arrows are shifted in the x-axis to make the plot clearer). The rightmost panel shows the age distribution for Terzan 5 (black histogram) compared to different age determinations for SOS1 (colored lines).}
    \label{fig:chemical-link}
\end{figure*}

Given that there are inherent uncertainties linked to the Galactic potential and other caveats (see Appendix \ref{app:orbits}), chemical analysis is needed to strengthen the above results further. SOS1 has a metallicity value that is in agreement with Terzan 5's most metal-poor population (PopC), which it is also the oldest one \citep{ferraro16}. {It is important to mention that the other survived GCs in our sample present a single metallicity distribution function, and all of them are more metal-rich than SOS1 (see Table \ref{tab:gcs_ref}).} Differently from a classic GC, Terzan 5 also shows populations of stars with different ages, a very old population ($>12$ Gyr) and an intermediate-age population ($\sim 5$ Gyr). Figure \ref{fig:chemical-link} shows SOS1's abundances of O, Mg, Ca, Si, C, and Al compared with Terzan 5 \citep{origlia11,origlia13}. Even though popC comprises only three stars, it is possible to observe good compatibility between SOS1 and popC for the $\alpha$-elements. 

The SOS1's chemical pattern, typical from a 2G GC, means that although most of the $\alpha$-elements will be preserved (especially Ca and Si), light elements will show variations with respect to the 1G stars in the cluster. The expected abundance differences of C, N, O, Mg, Al, and Si between 1G and 2G stars were provided by \cite{milone18}. Following their values, we estimate the expected variations from 1G to 2G stars for Terzan 5 most metal-poor PopC (Figure \ref{fig:chemical-link}; see Appendix \ref{app:age} for the details about the method employed to derive the 2G enrichments). The maximum possible abundance differences are indicated by the arrows in Figure \ref{fig:chemical-link}. It can be seen that the position of SOS1 in each of the [X/Fe] (for X = O, Mg, Ca, Si, C, and Al) is always within the range of the expected variations, confirming that SOS1 is chemically compatible with the PopC stars of Terzan 5. 

 Recently, \cite{romano23} derived the chemical evolution model for Terzan 5. They found that the cluster chemical pattern could be explained by a double-burst of a primordial cloud with a mass of $4\times10^7$ M$_{\odot}$ (with a delay of 5 Gyr between both bursts), and $\sim60\%$ of stars stripped. The authors also suggest that the metal-poor population of Terzan 5 is located mainly in the outer part of the cluster, being, therefore, easier to be stripped from the cluster and trapped by the Galactic bar than most internal stars, a scheme consistent with SOS1's properties. 

As mentioned before, the high Ce enrichment of SOS1 could indicate a very old age, therefore, an additional important link between SOS1 and Terzan 5 is the age. Although SOS1 has NASA K2 mission data \citep{Howell_2014}, the quality of the light curves was insufficient to allow the determination of its seismic age (see Appendix \ref{app:seismic}). Nevertheless, the other age determination method for red giants is based on the [C/N] abundance ratio, which is modified after the first dredge-up. This process brings new material synthesized in their cores through the CNO cycle toward the atmosphere. Since the maximum penetration during the first dredge-up depends on the stellar mass, the mixture brought up to the stellar surface also depends on the mass, which can be related to the stellar age \citep{masseron15,salaris15,lagarde17}. However, inside GCs in addition to the expected from stellar evolution, the 2G stars are also enriched in N and depleted by C because of the internal chemical evolution of the GC. Therefore, to use the [C/N]$-$age relations for 2G stars such as SOS1, the observed [C/Fe] and [N/Fe] need to be corrected by the expected enrichment and depletion typically happening in 2G stars in GCs. 

Applying the corrections derived above (represented by arrows in Figure \ref{fig:chemical-link}), we derived the age using the C/N chemical clock and adopting different calibrations (Appendix \ref{app:age}). Although the approach used here to estimate the age is very uncertain, this procedure can lead to, at last, a lower limit for the SOS1 star. We find an old age distribution for SOS1 and, therefore, compatible with the age distribution of Terzan 5 (rightmost panel of Figure \ref{fig:chemical-link}). 

\section{Conclusions}\label{sec:conclusions}
 
 {In this work, for the first time in our knowledge, we present an example of a chemodynamical tracing of a specific star not part of a known stellar stream back to an origin in a specific GC.} By means of the history of SOS1, we analyzed the influence of Terzan 5 on the formation of the Galactic bar in providing the field stellar population with N-Al-rich stars.

{We determine that SOS1 is likely a past member of a 2G population in a GC. Even though the origin from a completely destroyed GC remains as a possibility, we found a good chemodynamical agreement between SOS1 and Terzan 5, making Terzan 5 the most plausible survived GC from which SOS1 could be a member in the past. }

SOS1 exhibits distinct chemo-dynamical properties, including nitrogen enhancement ([N/Fe] = +0.71) and a significant aluminum enhancement ([Al/Fe] = +0.31), both of which are characteristic of 2G stars from GCs. These elements, along with its consistent orbit with the bar structure, strongly suggest that SOS1 was once part of a GC, {possibly} stripped from Terzan 5 after several close encounters with the Galactic bar at $-353\pm12(\pm107)$ Myr ago.

The metallicity of SOS1 strongly agrees with {the expected 2G stars of the most metal-poor population in Terzan 5}, popC, which has a median metallicity of [Fe/H]$\sim -0.80$. {The same chemical analysis was performed on the other GCs. Nevertheless, they have simple MDF and are more metal-rich than SOS1 (see Table \ref{tab:gcs_ref})}. Additionally, the SOS1 enhanced levels of s-process elements, such as cerium ([Ce/Fe] = +0.60), support the hypothesis that it retains chemical signatures from the early enrichment phases of the Milky Way’s inner regions. 

This study highlights SOS1 as a fossil record of the ancient processes that shaped the inner Galaxy. The chemical and orbital evidence converges to suggest that Terzan 5, and potentially other BFF, have significantly contributed to the build-up of the stellar populations in the Galactic bar. Investigating stars like SOS1 provides crucial insights into the role of GCs in the formation and evolution of the Milky Way bulge and bar regions. 

{This work thus provides strong foundation for similar studies in the future, with even better datasets coming from the next Gaia releases and large spectroscopic surveys.}


\begin{acknowledgments}
{The authors are grateful to the anonymous referee for his/her suggestions and remarks, which greatly improved the paper}. SOS and MV acknowledge the support of the Deutsche Forschungsgemeinschaft (DFG, project number: 428473034). SOS acknowledges a FAPESP PhD fellowship no. 2018/22044-3. SOS acknowledges the support from Nadine Neumayer's Lise Meitner grant from the Max Planck Society. SOS, MV, CC, JM, and DB thank the International Space Science Institute for supporting the ISSI team 542 on “CatS - A reference catalogue for Spectroscopic surveys” led by M. Valentini and G. Guiglion. A.P.-V. and S.O.S. acknowledge the DGAPA-PAPIIT grant IA103224. BB acknowledges partial financial support from FAPESP, CNPq and CAPES - Financial code 001. R.A.G. acknowledges the support from the PLATO Centre National D'{\'{E}}tudes Spatiales grant. D.B. acknowledges funding support by the Italian Ministerial Grant PRIN 2022, ``Radiative opacities for`astrophysical applications'', no. 2022NEXMP8, CUP C53D23001220006. J.M., D.B., and C.C. acknowledge support from the ERC Consolidator Grant funding scheme (project ASTEROCHRONOMETRY, G.A. no. 772293). Y.E. acknowledges the support of the UK Science and Technology Facilities Council (STFC). Part to the computations used in this paper were performed using the University of Birmingham's BlueBEAR HPC service, which provides a High Performance Computing service to the University's research community. See \url{http://www.birmingham.ac.uk/bear} for more details.
\end{acknowledgments}

\vspace{5mm}
\facilities{This paper includes data collected by the Kepler/K2 mission and obtained from the MAST data archive at the Space Telescope Science Institute (STScI). Funding for the Kepler/K2 mission is provided by the NASA Science Mission Directorate. STScI is operated by the Association of Universities for Research in Astronomy, Inc., under NASA contract NAS 5–26555.}

\software{The interactive figure was created using the \emph{plotly} python library (\url{https://plotly.com/python/}). The Galaxy image of Figure \ref{fig:main_plot} was created using the \emph{mw-plot} library available in \url{https://github.com/henrysky/milkyway_plot}.}


\appendix

\section{Interstellar reddening from DIBs}\label{app:dibs}

We obtained an independent measurement of extinction A$_V$ by measuring the equivalent width (EW) of near-infrared diffuse interstellar bands (DIBs) in the SDSS-IV/APOGEE spectrum of SOS1. Even though it is still unclear what creates DIBs, it has been widely shown that their features correlate with dust and gas in the interstellar medium. Hence, it is possible to measure A$_V$ via empirical relations \citep[e.g.][]{kos2014}. 

We used the DIB at $\lambda$ ~1.527 $\mu$m, using the same procedure as in \cite[e.g.][]{elyajouri2017}. The stellar normalised spectrum was divided by the best-fit model spectrum, and then we measured the EW of the DIB feature in the residual spectrum 
with a Gaussian fit. For better error estimation, we measured the EW 50 times, varying the continuum placement and limits for the fit each time. Line-of-sight extinction, A$_V$, has been then calculated using:
\begin{equation}\label{Eq:DIB}
EW_{DIB}= 102 {\rm m \mathring{A}} \times A_V^{1.01\pm0.01}
\end{equation}
The resulting reddening value is $A_{V}$ = 4.477 with $\sigma_{Av}$=0.351$_{\rm fit}$+0.564$_{\rm EW~meas}$ mag, where the associated error considers the error contribution from Eq.~\ref{Eq:DIB} and the EW measurement.


\section{Abundances verification}\label{app:abds_verify}

In order to validate the high values of N and Al provided by APOGEE for SOS1, we performed two different validation test:

\begin{enumerate}
\item We selected a star with similar atmospheric parameters gravity log~g, effective temperature T$_{\rm eff}$, metallicity [Fe/H], and SNR as SOS1, with solar N and Al. The reference star has APOGEE ID 2M07273047-7511343 (hereafter STARB, its information is in Table \ref{tab:informations}). Figure \ref{fig:spec_comp} shows the direct comparison between SOS1 and SARTB spectra. The upper panels compare the Al abundance for the lines  AlI $16718.90${\rm \AA\,} (left) and $16763.30${\rm \AA\,} (right). The blue line represents STARB, while SOS1 is in red. The direct comparison for N abundances is in the bottom panels. This analysis clearly shows the difference between both spectra for Al and N, showing that SOS1 is indeed N- Al-rich.

\begin{figure}[hbt!]
\centering
\includegraphics[width=0.8\textwidth]{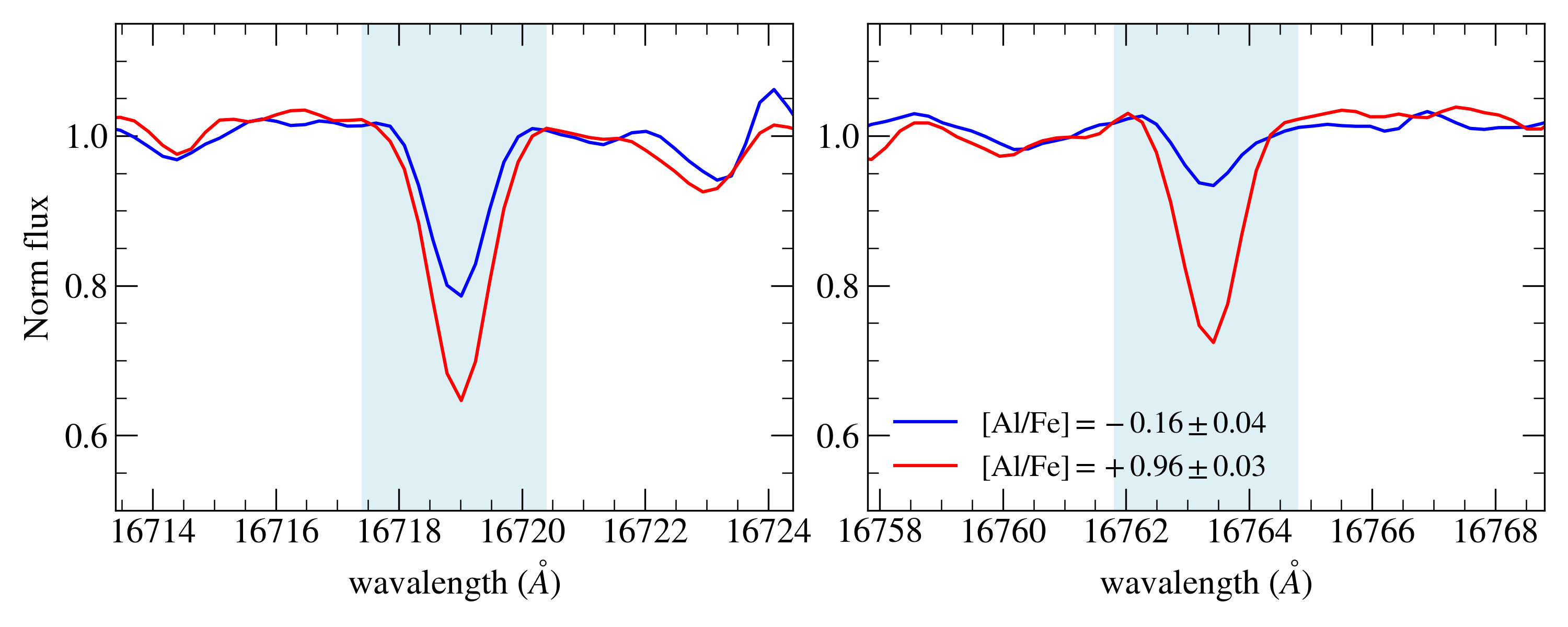}
\includegraphics[width=0.8\textwidth]{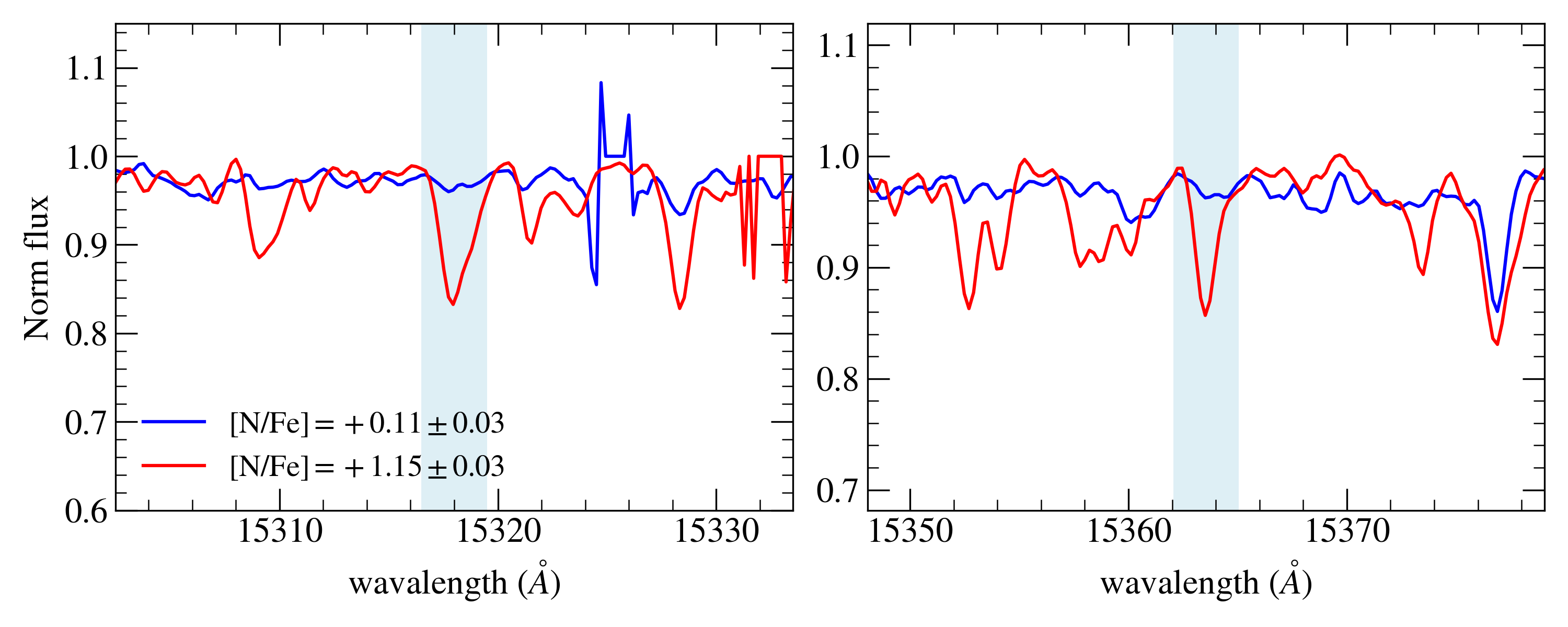}
\caption{Verification of the APOGEE-ASPCAP abundances of Al (upper panels) and N (bottom panels): comparison of spectra of the sample star SOS1 (red line) with a reference star (STARB) with similar stellar parameters and solar abundances (blue line). }
\label{fig:spec_comp}
\end{figure}

\item We remeasured Al, C, and N abundances through spectrum synthesis using the \texttt{MOOG2017} code \citep{sneden17} with solar abundances adopted from \cite{asplund09}. Figure \ref{fig:synthesis} shows the spectral fitting for the Al lines ALI $16718.90$ \AA\,\, (upper left) and $16763.30$ \AA\,\, (upper right), where the very high Al abundance is made clear from a comparison with the synthetic spectra for different [Al/Fe] values. For the C and N abundances (bottom panel), we overplotted synthetic spectra with different combinations of C, N, and O abundances in the wavelength between $15300.00$ \AA\,\, and $15400.00$ \AA\,\,. On average, the SOS1 observed spectrum is well reproduced by high N and low C values, as provided by APOGEE.

\begin{figure}[hbt!]
\centering
\includegraphics[width=0.39\textwidth]{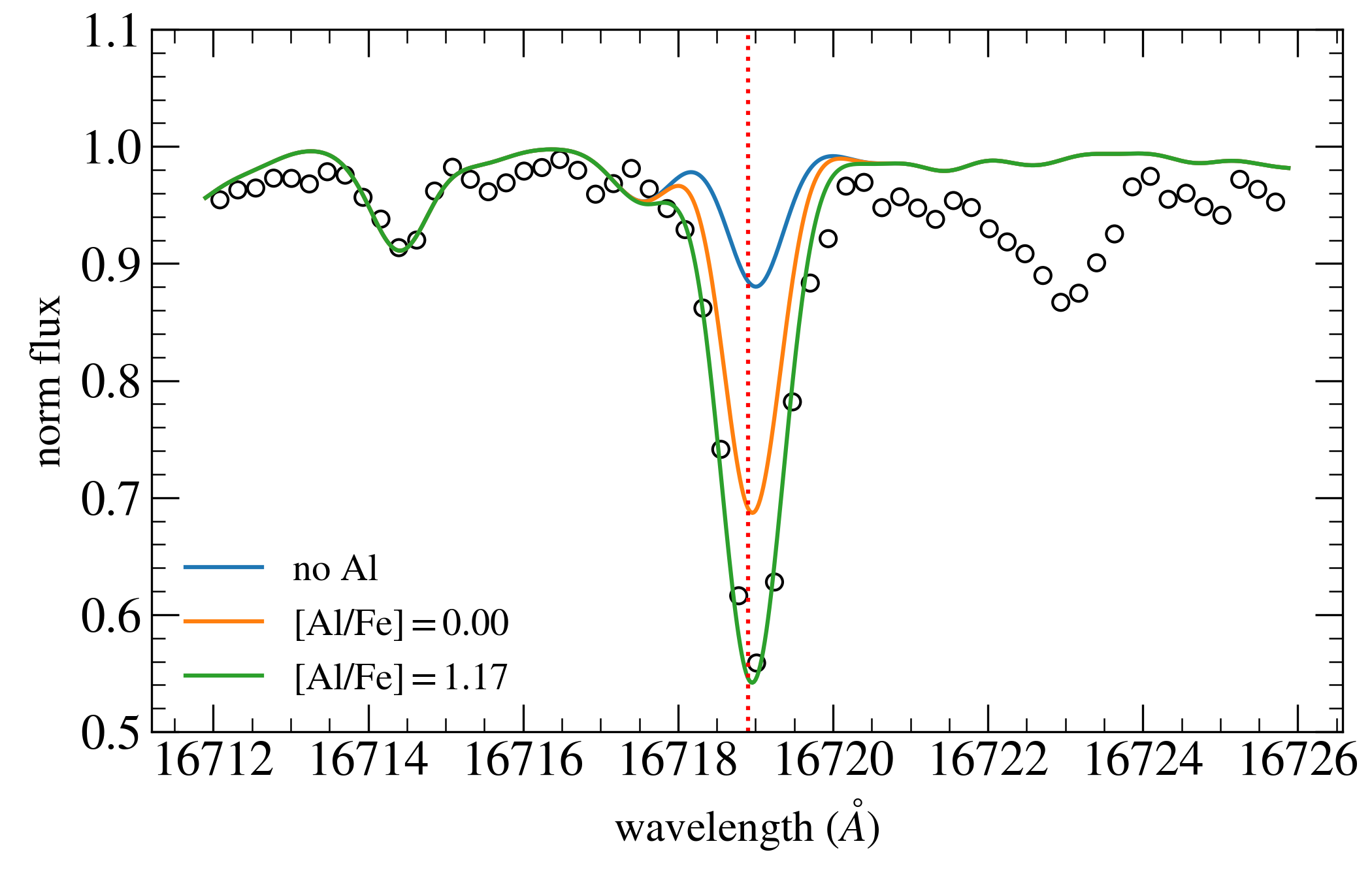}
\includegraphics[width=0.39\textwidth]{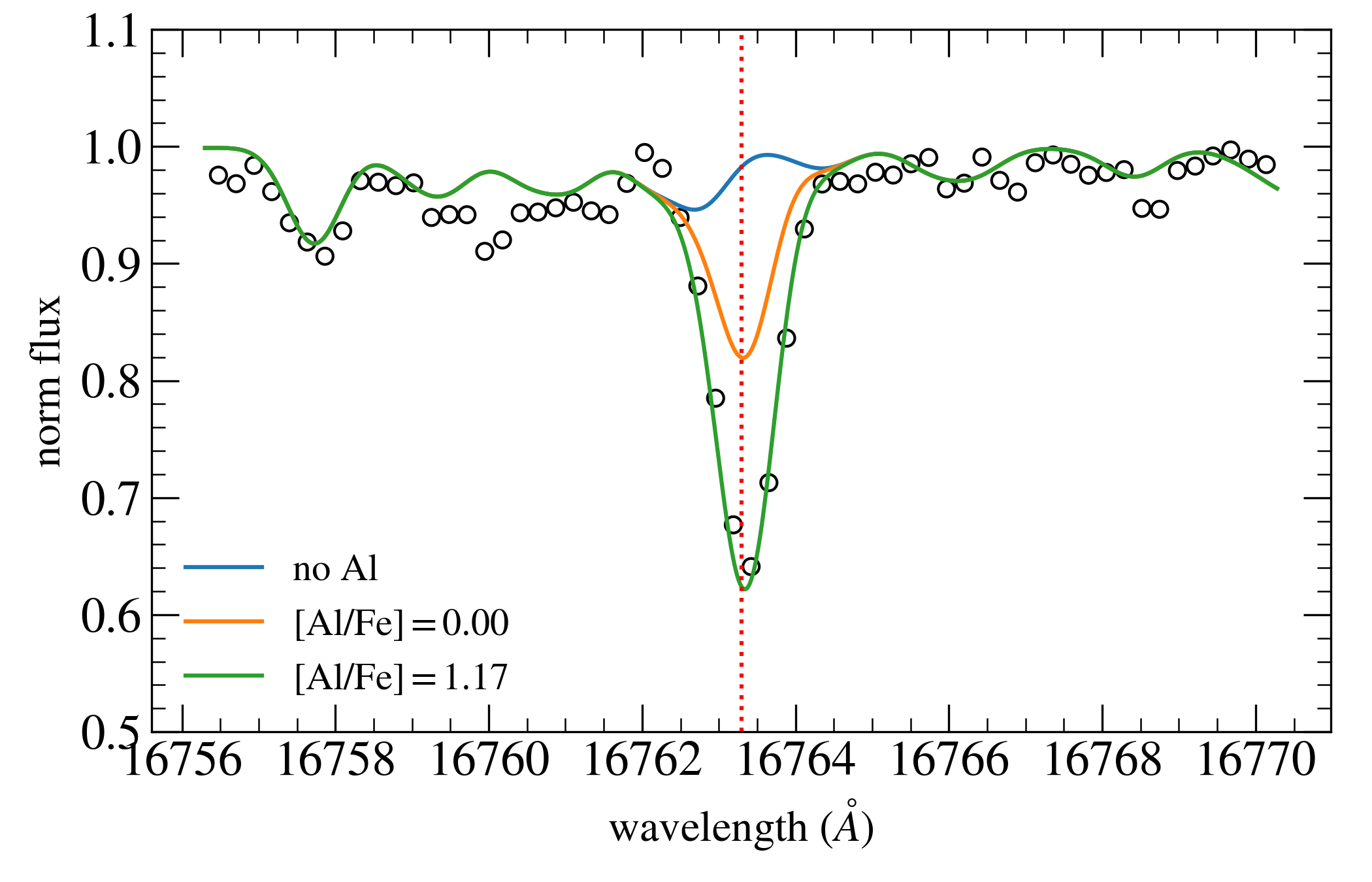}
\includegraphics[width=0.8\textwidth]{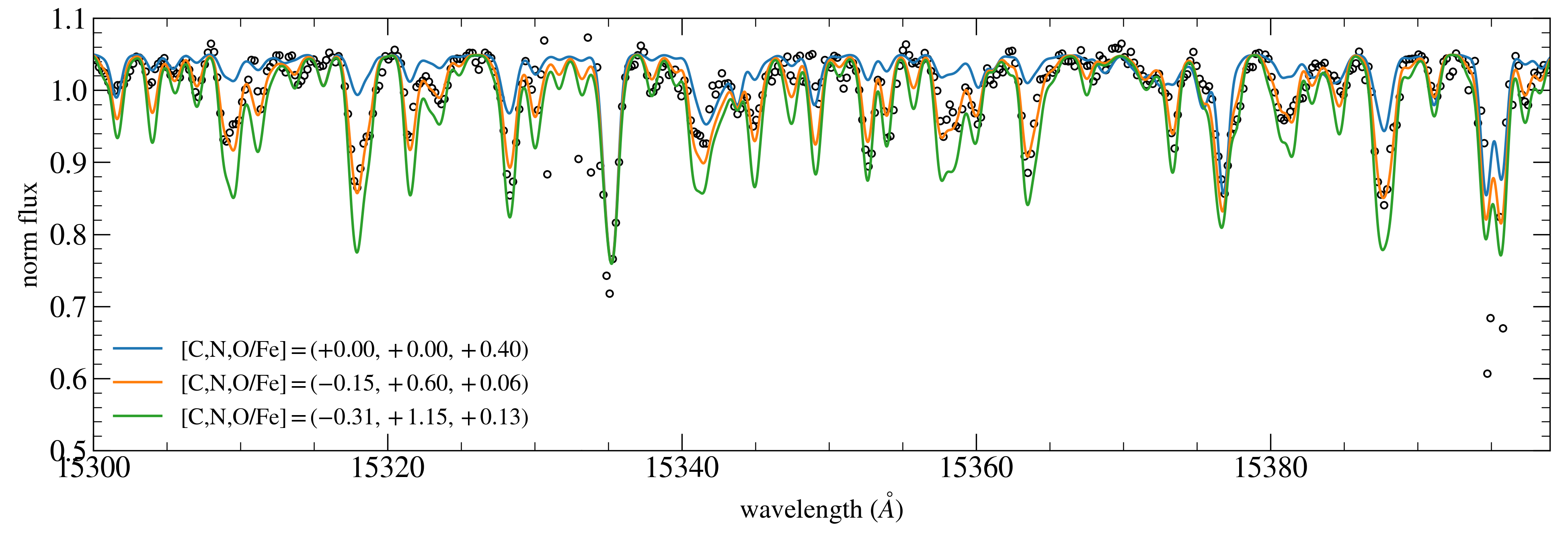}
\caption{Verification of APOGEE-ASPCAP abundances. Upper panels: Spectrum synthesis of Al. The coloured lines represent synthetic spectra with no Al abundance (blue), solar Al abundance (orange), and [Al/Fe]$=1.17$ (green, the value provided by APOGEE). Bottom panels: Spectrum synthesis of C, N, and O. The blue line is the normal CNO abundance set ${\rm [C, N, O/Fe]}=(0.0,0.0,+0.40)$, orange line ${\rm [C, N, O/Fe]}=(-0.15,+0.60,+0.06)$, and green line ${\rm [C, N, O/Fe]}=(-0.31,+1.15,+0.13)$ the set provided by APOGEE.}
\label{fig:synthesis}
\end{figure}

\end{enumerate}

 These qualitative validation tests show that the abundances provided by APOGEE DR17 for SOS1 (and STARB) are not due to artifacts in the \texttt{ASPCAP} pipeline.

\section{Orbital analysis}\label{app:orbits}

For each object, SOS1 and all 165 MW GCs, we generated 500 initial conditions through a Monte Carlo (MC) resampling considering the errors in proper motions, radial velocity, and distance. 

We employed well-known orbital classification criteria by \cite{portail15} to compute the probability of the stars that support the bar structure. To do that, for each orbit, we computed the fast Fourier transform to identify the main frequencies in the Cartesian x and z-coordinates and the Cylindrical radius R. With these frequencies, we can analyse the ratios $f_R/f_X$ and $f_Z/f_X$, which provide a numerical interpretation of how many radial or vertical excursions the orbits have for one loop, respectively. Orbits that support the bar structure must be confined no farther than the corotation radius, which for the adopted model is placed at $\sim6$ kpc, and also has $f_R/f_X = 2.0\pm0.1$  \citep{portail15}. Therefore, the final probability of following the bar ($\mathcal{P}_{\rm Bar}$) for the object is the number of orbits with $f_R/f_X = 2.0\pm0.1$ divided by the total of the initial conditions, in this case, 500.

Since SOS1 is not a hypervelocity star, the hypothesis of smooth capture is the most probable scenario to explain its current configuration: runaway star member of a GC 2G population. It is worth mentioning that another possible explanation is that the SOS1 parent cluster was completely dissolved in the inner part of the Galaxy. This scenario needs a Galactic potential, which evolves with time and formalism for tidal forces and dynamical friction. For that reason, we will not explore this scenario in this work. We selected the GCs with E$_J$ compatible within $1\sigma$ with SOS1 E$_J$ value. The left plot on panel c of Figure \ref{fig:main_plot} shows the E$_J$ for the most compatible GCs. We show the input and output information of orbital integration, and the probability of belonging to the bar for the most likely SOS1 parent cluster in Table \ref{tab:gcs_ref}. The observational parameters of the GCs were taken from Baumgardt's compilation \footnote{\url{https://people.smp.uq.edu.au/HolgerBaumgardt/globular/}}.

\begin{table*}[htb!]
    \centering
    \caption{\textbf{Orbital input and output parameters of SOS1 and its most likely parent cluster}. The input values were taken from Baumgardt's compilation$^1$, and the output was obtained from orbital integration.}
    \label{tab:gcs_ref}
    \resizebox{\textwidth}{!}{%
    \begin{tabular}{l l||c|c|c|c|c|c|c}
    \hline
    \noalign{\smallskip}
        & & SOS~1 & Palomar~1 & Terzan~2 & Terzan~5 & NGC~6440 & Terzan~6 & NGC~6528 \\     
    \noalign{\smallskip}
    \hline
         \hline
    \noalign{\smallskip}
        & & \multicolumn{7}{c}{Input parameters}\\ 
    \noalign{\smallskip}
        \hline
        RA  & (deg)                         & $266.446$       & $+53.334$      & $261.888$      & $267.020$ & $267.220$ & $267.693$ & $271.207$ \\ 
        DEC & (deg)                         & $-26.653$       & $+79.581$      & $-30.802$      & $-24.779$ & $-20.360$ & $-31.275$ & $-30.056$ \\ 
        R$_{\rm sun}$ &(kpc)                & $6.9\pm1.0$     & $11.1\pm0.3$   & $7.8\pm0.3$    & $6.6\pm0.2$ & $8.3\pm0.2$ & $7.3\pm0.4$ & $7.8\pm0.2$ \\ 
        $<$RV$>$ &(km s$^{-1}$)             & $-75.4\pm0.1$   & $-75.7\pm0.2$  & $133.5\pm0.7$  & $-81.9\pm0.9$ & $-69.4\pm0.9$ & $137.2\pm1.7$ & $211.9\pm0.4$ \\ 
        $\mu_{\alpha}^{*}$ &(mas yr$^{-1}$) & $-1.45\pm0.02$  & $-0.26\pm0.03$ & $-2.14\pm0.03$ & $-1.86\pm0.03$ & $-1.19\pm0.02$ & $-4.99\pm0.05$ & $-2.17\pm0.02$ \\ 
        $\mu_{\delta}$&(mas yr$^{-1}$)      & $-6.20\pm0.06$  & $+0.02\pm0.03$ & $-6.26\pm0.03$ & $-5.11\pm0.03$ & $-4.00\pm0.02$ & $-7.46\pm0.05$ & $-5.64\pm0.02$ \\ 
        Mass &(M$_\odot$)                   & ---             & $(9.26\pm1.83)\times10^2$ & $(8.05\pm2.31)\times10^4$ & $(1.09\pm0.08)\times10^6$ & $(5.69\pm0.45)\times10^5$ & $(1.00\pm0.01)\times10^5$ & $(9.44\pm0.91)\times10^4$ \\ 
        Mass$_{\rm ini}$ &(M$_\odot$)       & ---             & $1.23\times10^4$ & $5.89\times10^6$ & $4.20\times10^7$ & $6.17\times10^6$ & $4.57\times10^6$ & $2.14\times10^6$ \\ 
        r$_t$ &(pc)                         & ---             & $22.84$ & $12.45$ & $51.26$ & $33.88$ & $13.89$ & $10.98$ \\ 
        Age   &(Gyr)                        & $>10.0$         & $7.3\pm1.2$ & --- & $12.0\pm1.0$ & $13.0\pm1.5$ & --- & $11.0\pm1.0$ \\ 
        $\rm [Fe/H]$ &                            & $-0.96\pm0.06$  & $-0.65\pm0.10$ & $-0.69\pm0.10$ & $-0.23\pm0.10$ & $-0.36\pm0.10$ & $-0.56\pm0.10$ & $-0.11\pm0.10$ \\
        $\rm [Fe/H]$ FLAG &  (\# of peaks)         & ---             & 1              &      1         &  3             &      1         &    1           & 1 \\
        \hline
    \noalign{\smallskip}
        & & \multicolumn{7}{c}{Output parameters}\\ 
    \noalign{\smallskip}
        \hline
        r$_{\rm apo}$ &(kpc) & $1.6\pm0.3$  & $21.6\pm0.4$ & $0.9 \pm0.1$ & $1.9 \pm0.1$ & $1.5 \pm0.2$ & $1.5 \pm0.2$ & $1.7 \pm0.2$ \\ 
        r$_{\rm peri}$ &(kpc)& $0.04\pm0.02$ & $15.81\pm0.14$ & $0.21\pm0.06$ & $0.10\pm0.02$ & $0.04\pm0.02$ & $0.03\pm0.01$ & $0.04\pm0.02$ \\ 
        $|Z|_{\rm zm }$ &(kpc)& $0.71\pm0.20$ & $5.51\pm0.14$ & $0.49\pm0.08$ & $0.87\pm0.06$ & $1.23\pm0.05$ & $1.01\pm0.09$ & $1.11\pm0.06$ \\ 
        $ecc$ & & $0.95\pm0.02$ & $0.16\pm0.01$  & $0.63\pm0.12$ & $0.89\pm0.02$ & $0.95\pm0.03$ & $0.96\pm0.02$ & $0.96\pm0.02$ \\ 
        E$_J$ & ($10^5$km$^2$ s$^{-1}$)& $-1.652\pm0.072$ & $-1.635\pm0.008$ & $-1.699\pm0.018$ & $-1.602\pm0.009$ & $-1.582\pm0.007$ & $-1.620\pm0.032$ & $-1.582\pm0.008$ \\ 
        $\mathcal{P}_{\rm bar}$ & $\%$ & $84$ & $0$ & $0$ & $94$ & $0$ & $9$ & $27$ \\ 
        $\mathcal{N}_{\rm diss}$ & (\# of points) & --- & $0$ & $0$ & $3794$ & $15$ & $1$ & $0$ \\ 
        $\mathcal{N}_{\rm diss}> -2$ Gyr & (\# of points) & --- & $0$ & $0$ & $936$ & $0$ & $0$ & $0$ \\ 
        \hline
\end{tabular}}
\end{table*}

With the selection of the SOS1 most likely parent clusters, as mentioned before, mainly from dynamics, we compared SOS1 and each GC, considering the entire orbit. To do that, we compared each orbit of SOS1 with each GC,  in total, for each pair SOS1-GC, $500 \cdot 500 = 2.5\cdot10^5$ combinations. 

As the first step, we computed the point of the orbit where SOS1 is within the GC tidal radius. We employed literature values of tidal radius \citep{baumgardt22}. During the orbital time, SOS1 can move through all the MW GCs during its orbits. Therefore, we only accepted the close encounter point where the star is also gravitationally bound to the cluster potential. We calculated the total energy of SOS1 inside the GC for all close encounter points due to the cluster potential, where the points with negative energy were considered gravitationally bound. For the combinations with more than one close encounter gravitationally bound point, we considered only the first one (closest to the present day) and assumed that SOS1 belongs to the cluster before the first point. The number of combinations with gravitationally bound close encounter points shows the probability of  SOS1 being a cluster member. Also, it provides the dissociation time when SOS1 was completely captured from the cluster by the Galactic bar. For the selected GCs (see Figure \ref{fig:main_plot}), only Terzan 5 has substantial close encounter points for the $2.5\cdot10^5$ orbit combinations (Table \ref{tab:gcs_ref}).

\section{Note on the SOS1 light-curve}\label{app:seismic}
Based on the atmospheric parameters from APOGEE-DR17 and scaling relations \citep{kjeldsen95}, we expect SOS1 to exhibit a power spectrum with an excess due to solar-type oscillations at approximately $4\mu$Hz. Unfortunately, several factors have made the extraction of the solar oscillation signal impossible. Firstly, this campaign was interrupted due to a configuration error in the drift control, resulting in a reduced duty cycle and a difference in signal levels between the two portions of the light curve. Additionally, this star was located in CCD2.1 of the \emph{Kepler}/K2 field. For this campaign, which focused on the bulge direction, CCD2.1 is characterized by a high density of sources, more than three orders of magnitude higher than a typical \emph{Kepler}/K2 field. A test conducted by degrading a \emph{Kepler} target light curve with similar log(g) and $T_{\rm eff}$ to K2-like characteristics indicates that the noise level in SOS1's light curve is a thousand times higher than in the simulated one. This suggests that the high noise level is what makes a robust detection of solar-type oscillations impossible.

\section{Age derivation}\label{app:age}

We attempted to constrain the stellar age following two different approaches.

\subsection{Bayesian Age determination}

The mass and age of SOS1 were derived using the Bayesian code PARAM\footnote{\url{http://stev.oapd.inaf.it/cgi-bin/param}} \citep[][and references therein]{PARAM3}, which matches observed quantities ($T_{\rm eff}$, [Fe/H], luminosity) to a grid of stellar evolutionary tracks. The unreliable \emph{Gaia} parallax due to SOS1's distance was replaced by the Starhorse \citep{queiroz23} distance (Table \ref{tab:informations}).

We computed three grids of models using the MESA code \citep[r-11701][]{paxton18}, covering a mass range of 0.6 to 2.5 M$\odot$ with 0.05 M$\odot$ steps. The grids differ in alpha-element enhancement: solar, ${\rm[\alpha/Fe]}=0.2$, and ${\rm[\alpha/Fe]}=0.4$, with 12 iron content values from ${\rm[Fe/H]}=-2.50$ to $0.25$ \citep{asplund09}. Opacity tables were computed using OPAL for the stellar interior and Wichita University data (private communication) for low temperatures.
The atmospheric boundary conditions follow the KS grey $T(\tau)$ law, and convection is modelled using MLT \citep{cox68}. A diffusive model for extra mixing at the convective envelope base \citep{kahn19} and core convection in the HeB phase \citep{bossini17} were applied. Microscopic diffusion was also included.

Models were followed from the PMS phase to the first AGB pulse or peak luminosity post-He burning. Radial mode oscillations were computed for models along evolutionary tracks to sample different phases. These models form the basis of PARAM, which estimates stellar parameters using $T_{\rm eff}$, log g, ${\rm [Fe/H]}$, ${\rm [\alpha/Fe]}$, and seismic parameters \citep{rodrigues17}.

The output provides the probability distributions of mass, age, radius, and density. SOS1's broad age distribution (rightmost panel of Figure \ref{fig:chemical-link}) arises from degeneracies in red giant evolutionary tracks, making it hard to distinguish between different masses and ages without seismology.

\subsection{Chemical clock: [C/N]--age relation}

The [C/N]$-$age relation is widely used in the literature. The main idea is that, for red giants, the abundance of N and C in the stellar surface changes during the evolution after the first dredge-up. The variation depends on the mass and, consequently, the age. Before applying the calibrations, we remove the effect of MPs from the C and N abundances of SOS1. This correction is necessary because the calibrations derived in the literature \citep[e.g.][]{casali19,spoo22} are based purely on stellar evolution. The corrections are as follows:

\begin{equation}
    [{\rm C/N}]_{\rm SOS1} = [{\rm C/N}]_{\rm 1G} + [{\rm C/N}]_{\rm MPs}
\end{equation}
basically, the abundances derived from APOGEE for 2G stars are the abundance of 1G stars added by a contribution from MPs: enhancement for N and depletion for C. The variations due to MPs were derived using the values considering all GCs and doing a linear regression for logMass \citep{baumgardt22} vs. abundance variations \citep{milone18}. The differences are then obtained by interpolating for the Terzan 5's mass \citep{baumgardt22}.

\begin{figure}[hbt!]
    \centering
    \includegraphics[width=0.8\textwidth]{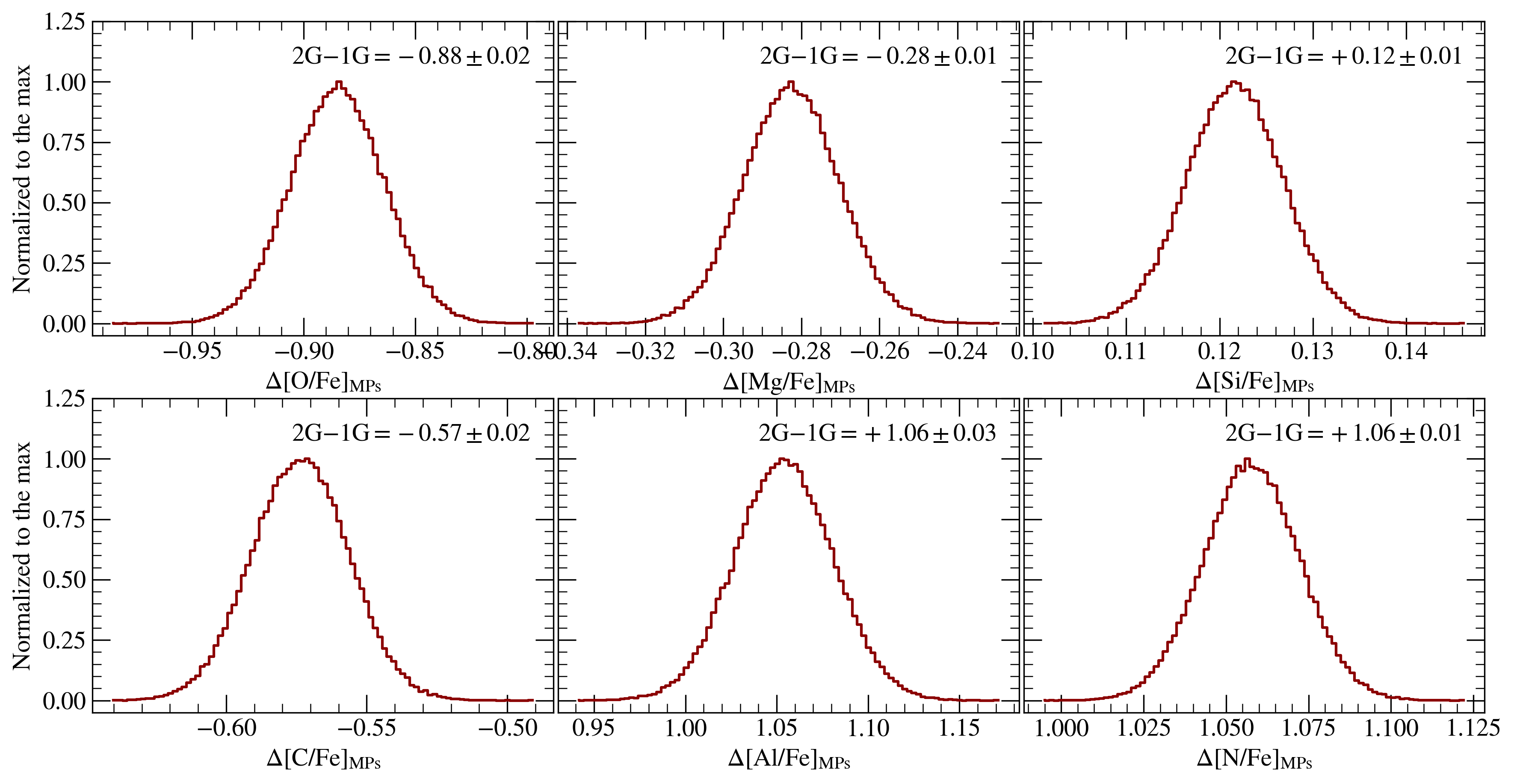}
    \caption{{Derivation of the abundance contribution for the elements O, Mg, Si, C, Al, and N due to the MPs.}}
    \label{fig:deltas}
\end{figure}

The values are represented in Figure \ref{fig:chemical-link} as red arrows. We placed the arrows starting from the median value of the Terzan 5's popC since \cite{origlia13} have not found evidence of 2G stars in their sample. It is also important to note that the N abundance is not presented in Figure \ref{fig:chemical-link} because there are no calculations for this element. The variations due to MPs are then removed (or added) from the values derived by APOGEE, resulting in the possible 1G values. The distributions are displayed in Figure \ref{fig:deltas}.

Having obtained the corrected [C/N] abundance, we then applied the [C/N]$-$age relation using four different calibrations: \cite{spoo22} for APOGEE DR16 and DR17; \cite{casali19} derived using \textit{Gaia}-ESO Survey. The rightmost panel of Figure \ref{fig:chemical-link} shows the resulting age distributions for each calibration. For comparison purposes, we also plotted the age distribution for Terzan 5 \citep{ferraro16}.

\bibliography{sample631}{}
\bibliographystyle{aasjournal}

\end{document}